\documentclass[letterpaper, 10pt, conference]{ieeeconf} 
\IEEEoverridecommandlockouts
\usepackage{graphicx}
\usepackage{xcolor}
\usepackage{amsmath}
\usepackage{mathtools}
 \mathtoolsset{showonlyrefs}
\usepackage{tikz}
\usepackage{verbatim}
\usepackage{subcaption}
\usepackage{pgfplots}
\pgfplotsset{compat=1.10}
\usepackage{epstopdf} 
\usepackage{relsize}
\usepackage{booktabs}
\definecolor{Gray}{gray}{0.90}
\usepackage{colortbl}

\usepackage{algorithm}
\usepackage{algpseudocode}
\usepackage{color}
\usepackage{blkarray}
\usepackage{sidecap}
\usepackage{multirow}

\usepackage{amssymb}
\usepackage{url}

\newcommand{\calN}[0]{\mathcal{N}}

\definecolor{amber}{rgb}{1.0, 0.75, 0.0}
\newcommand\figref{Fig.~\ref}

\usepackage{xurl}
\usepackage{hyperref}
\usepackage{resizegather}

\usepackage{setspace}

\usepackage{makecell}
\usepackage{censor}

\allowdisplaybreaks
\usepackage{stfloats}

\usepackage{eso-pic}
\newif\ifarxiv
\arxivtrue

\begin{document}{

\title{\bf \texorpdfstring{DM$^3$-Nav}{DM3-Nav}:  Decentralized Multi-Agent Multimodal Multi-Object Semantic Navigation}
\author{ Amin Kashiri, Atharva Jamsandekar, Yasin Yaz{\i}c{\i}o\u{g}lu
\thanks{A. Kashiri is with the Department of Electrical and Computer Engineering at Northeastern University, Boston, MA ({\tt kashiri.a@northeastern.edu}).}%
\thanks{A. Jamsandekar is with the Department of Electrical and Computer Engineering at Northeastern University, Boston, MA ({\tt jamsandekar.a@northeastern.edu}).}%
\thanks{Y. Yaz{\i}c{\i}o\u{g}lu is with the Departments of Mechanical and Industrial Engineering and Electrical and Computer Engineering at Northeastern University, Boston, MA  ({\tt y.yazicioglu@northeastern.edu}).}%
}
}

\ifarxiv
\AddToShipoutPictureBG*{%
  \AtPageUpperLeft{%
    \put(\LenToUnit{0.5\paperwidth},\LenToUnit{-0.45in}){%
      \makebox[0pt][c]{\footnotesize Accepted to the 2026 IEEE/RSJ International
        Conference on Intelligent Robots and Systems (IROS 2026)}}}%
  \AtPageLowerLeft{%
    \put(\LenToUnit{0.5\paperwidth},\LenToUnit{0.4in}){%
      \makebox[0pt][c]{\parbox{0.8\paperwidth}{\centering\scriptsize
        \copyright\ 20XX IEEE.  Personal use of this material is permitted.  Permission from IEEE must be obtained for all other uses, in any current or future media, including reprinting/republishing this material for advertising or promotional purposes, creating new collective works, for resale or redistribution to servers or lists, or reuse of any copyrighted component of this work in other works.
      }}}}%
}
\fi

\maketitle
\overrideIEEEmargins   
\begin{abstract}

We present \texorpdfstring{DM$^3$-Nav}{DM3-Nav}, a fully decentralized multi-agent semantic navigation system supporting multimodal open-vocabulary goal specification and multi-object missions. In our setting, decentralization implies operation without a central coordinator, global map aggregation, or shared global state at runtime. Robots operate autonomously and coordinate through ad-hoc pairwise communication, exchanging local maps, goal status, and navigation intent without synchronization. An implicit task allocation mechanism combining intent broadcasting and distance-weighted frontier selection reduces redundant exploration while preserving decentralized operation. Evaluations on HM3DSem scenes using the HM3Dv0.2 and GOAT-Bench datasets demonstrate that \texorpdfstring{DM$^3$-Nav}{DM3-Nav} matches or exceeds centralized and shared-map baselines while eliminating single points of failure inherent in centralized architectures. Finally, we validate our approach on two mobile robots in a real-world office, relying entirely on onboard sensing and computation. Code and a video of our real-world experiment are available online.\footnote{\url{https://github.com/Multi-Agent-Robotics-and-Autonomy-Lab/DM3_Nav}}

\end{abstract}

\section{Introduction}
\label{intro}

Multi-robot systems offer significant potential for search, inspection, and retrieval tasks that require locating multiple objects specified through diverse modalities, including category labels, language descriptions, or reference images. Achieving robust and scalable multi-object semantic navigation with robot teams in unknown environments requires coordination without centralized planning or shared global state. However, existing approaches address these requirements only partially: some focus on multimodal or multi-object navigation but are limited to single-agent settings, while others introduce multi-agent coordination but rely on centralization and support only categorical goals.

Early work on object-goal navigation introduced semantic maps for category-level navigation \cite{sem_exp2020}. Gadre et al. \cite{cow2023} showed that CLIP-based representations enable zero-shot navigation to object categories, and Yokoyama et al. \cite{vlfm2024} grounded vision-language semantics into persistent frontier maps for long-horizon planning. However, these approaches address only a single goal per episode. Multi-object navigation (MON) extends this setting by requiring a sequence of objects to be located within an episode. The MultiON benchmark \cite{multion2020} formalized this task. Subsequent approaches addressed MON through reinforcement learning \cite{seqagnostic2023, zeng2023multi}, but require training and support only categorical goals. More recently, zero-shot object navigation (ZSON) methods have emerged. OneMap \cite{onemap2025} builds reusable open-vocabulary semantic maps, and WTRP-Searcher \cite{tsp_zson2025} formulates frontier selection as a weighted traveling repairman problem. GOAT \cite{goat2023} further supports multimodal goals specified through language and images via instance-aware semantic maps. While these methods support open-vocabulary queries and multi-object sequences, they remain limited to single-agent operation.

In multi-agent settings, Co-NavGPT \cite{conav2023} uses a centralized vision-language planner operating on a shared map. MCoCo-Nav \cite{mcoconav2025} performs decentralized reasoning but relies on a globally aggregated semantic map, introducing implicit centralization. Moreover, both systems support only single categorical goals. In the related domain of multi-agent exploration, the literature provides foundations for decentralized coordination. Frontier-based approaches achieve coordination through map-sharing \cite{multi_frontier1998}, while RACER \cite{racer2023} and related methods \cite{frontier_nbv2024} achieve fully decentralized exploration via efficient map representations and pairwise interactions. A complementary line of work studies collaborative SLAM, where the goal is to build globally consistent maps and trajectories. Centralized approaches such as CCM-SLAM \cite{2019ccm} and COVINS \cite{2021covins} aggregate data at a central server, while decentralized alternatives such as DOOR-SLAM \cite{2020door}, Kimera-Multi \cite{2022kimera}, and Swarm-SLAM \cite{2023swarmslam} coordinate through peer-to-peer communication, using distributed pose graph optimization and inter-robot loop closure detection. Multi-robot active SLAM further couples mapping with exploration, using information-theoretic or frontier-based methods \cite{2023activeslam, burgard2005coordinated} to jointly reduce map uncertainty. While these methods provide effective solutions for multi-robot exploration and SLAM, they do not address decentralized object search or semantic navigation.


\begin{figure*}[t]
    \centering
    \includegraphics[width=0.85\textwidth]{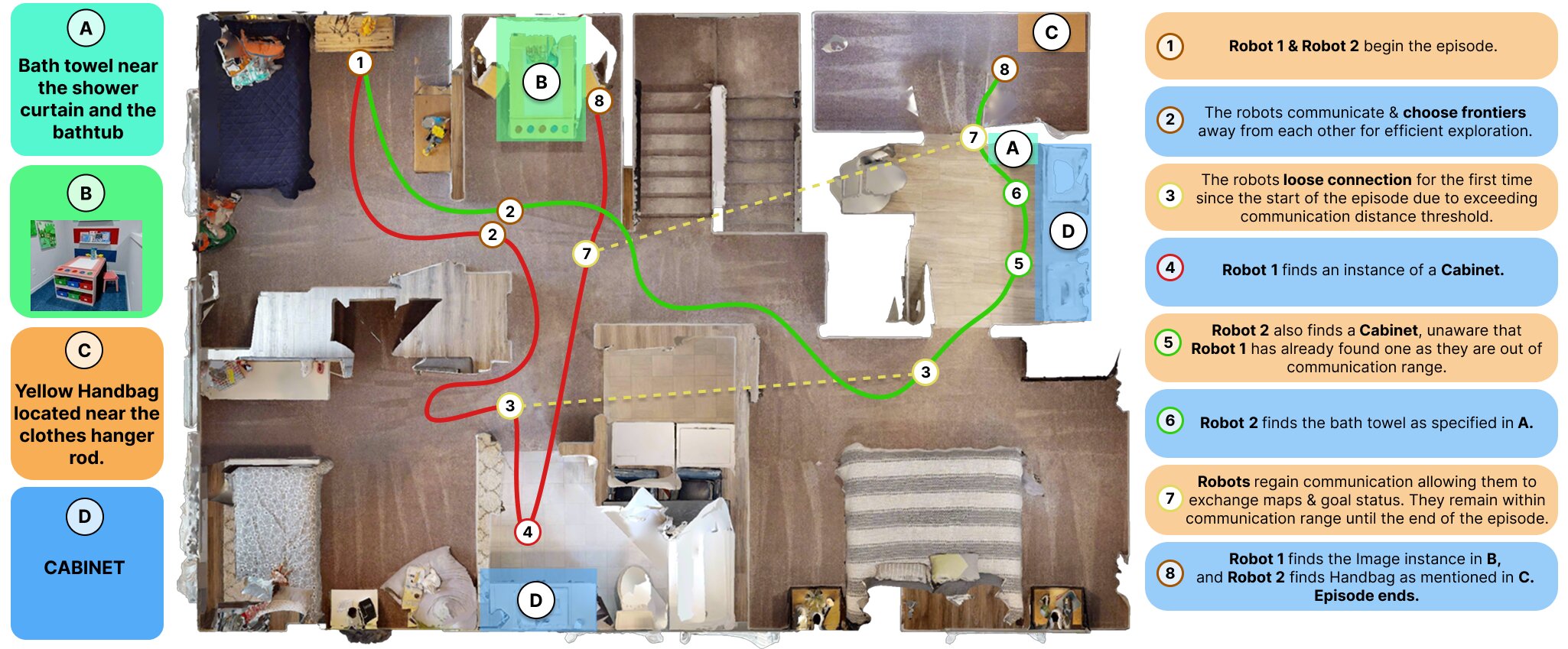}
    \caption{Overview of a multi-agent multimodal semantic navigation episode. Two robots (red and green paths) explore an unseen environment to successfully locate four distinct targets: (A) towel [language], (B) table [image], (C) handbag [language], (D) cabinet [category].}
    \label{fig:visual}
    \vspace{-10pt}
\end{figure*}

We present, to the best of our knowledge, the first multi-agent semantic navigation system that jointly supports multimodal goals (category, language, and image) and multi-object episodes under fully decentralized coordination. Our approach employs an implicit coordination mechanism that combines intent broadcasting and distance-weighted frontier selection to achieve task allocation without explicit negotiation, while each robot leverages an instance-aware semantic mapping framework based on GOAT \cite{goat2023}. In particular, our main contributions are as follows:
\begin{itemize}
\item A fully decentralized multi-agent semantic navigation architecture that eliminates centralized planning and shared global maps while enabling multimodal goal specifications and multi-object episodes.
\item Multi-object Multi-agent SPL (MSPL), a metric that extends SPL to settings with multiple robots, multiple objects, and multiple valid instances per object by measuring efficiency relative to the optimal makespan.
\item Extensive evaluation on HM3DSem scenes \cite{hm3dsem2023} using HM3Dv0.2 \cite{habitatchallenge2023} and GOAT-Bench \cite{goatbench2024}, demonstrating that the proposed decentralized approach matches or exceeds centralized and shared-map baselines.
\item Real-world deployment on a multi-robot platform validating the proposed approach in practice.
\end{itemize}
We additionally identify and address several limitations in the GOAT navigation pipeline that yield substantial performance improvements, even in the single-agent case (Appendix~\ref{appendix:modifications}).

\section{Problem Formulation}
\label{prob_formulation}

We consider a decentralized multi-agent semantic navigation task where a team of $n$ robots must cooperatively locate a set of target objects in an unknown environment.

\textbf{Task Definition.} An episode $E = (S, P, G)$ consists of a scene $S$, initial robot positions $P = \{p_1, \ldots, p_n\}$, and a set of goal specifications $G = \{g_1, \ldots, g_m\}$. Each goal $g_j$ can be specified in one of three modalities: (1) a category label (e.g., ``tv''), (2) a natural language description (e.g., ``the black table next to the door''), or (3) a reference image of the goal object. Robots have no prior knowledge of the environment and may find the goals in any order.

\textbf{Observations and Actions.} At each timestep $t$, robot $i$ receives an observation $o_t^i = (I_t^i, D_t^i, p_t^i)$ consisting of an RGB image, depth image, and the robot's pose. Based on \(o_t^i\) and its internal state, the robot selects a navigation action $a_t^i$.

\textbf{Decentralization Constraints.} Robots operate without access to a centralized server or shared global map. Each robot maintains its own local representation of the environment and coordinates through ad-hoc pairwise exchange of local maps, goal status, and navigation intent when two robots are in communication.

\textbf{Success Criteria.} A goal $g_j$ is found when any robot stops within a threshold distance of a matching object. Robots operate under a maximum step budget, after which the episode terminates regardless of how many goals were found. 

\textbf{Objective.} The objective is to maximize goals found while minimizing total navigation cost. We formalize this as Multi-agent SPL (MSPL) in Section~\ref{subsec:eval_metrics}, which combines success rate with the ratio of optimal to actual path length.

\section{Methodology}
\label{method}
We present a decentralized multi-agent navigation system for locating multiple goals specified through category labels, language descriptions, or images. \figref{fig:visual} shows two robots coordinating to locate four such targets in an indoor environment. By exchanging information when in range, the agents achieve scalable exploration and task allocation without relying on a centralized planner or continuous connectivity. \figref{fig:overview} provides an overview of our architecture.

\begin{figure*}[t]
    \centering
    \includegraphics[width=0.85\textwidth]{DM3_Flowchart.png}
    \caption{Overview of the \texorpdfstring{DM$^3$-Nav}{DM3-Nav} architecture. Each robot operates autonomously with its own perception, memory, and planning modules. Decentralized coordination is achieved by exchanging semantic maps, goal status, and navigation intent through local communications with in-range robots.}
    \label{fig:overview}
    \vspace{-10pt}
\end{figure*}
\subsection{System Overview}
Each robot independently processes RGB-D observations to build a spatial-semantic map, detect and localize objects, and plan navigation. 
Unlike multi-agent methods that require a shared global map \cite{mcoconav2025} or central coordination \cite{conav2023}, our approach coordinates through ad-hoc pairwise information exchange. When two robots can communicate, they share map information and goal status, enabling implicit task allocation without global synchronization. Because coordination is local and pairwise, each robot's computational and communication overhead depends only on its neighborhood, not the total team size, so the system scales naturally.

\subsection{Single-Agent Architecture}
For individual robot perception, mapping, and goal matching, we adopt the modular architecture from GOAT \cite{goat2023}. Each robot maintains a 2D top-down semantic map with geometric channels (obstacle, explored, navigable) and semantic channels for detected instances. An instance memory $\mathcal{I}$ stores multi-view appearance observations for each detected object, enabling robust goal matching across viewpoints. Goals are matched by category lookup, CLIP-based \cite{clip2021} similarity for language, or SuperGlue \cite{superglue2020} feature matching for images. The robot navigates to matched instances or explores frontiers until a match is found. Modifications to the original GOAT design are detailed in Appendix~\ref{appendix:modifications}.

\subsection{Decentralized Coordination}
\label{subsec:dec_coord}
\subsubsection{Communication Protocol}
Coordinating robots without a central planner requires mechanisms for information sharing and implicit task allocation, which we achieve through local pairwise opportunistic communication. 
To reduce communication overhead, each robot maintains a per-neighbor cooldown period $\tau$: after sending a full map to robot $r_j$, robot $r_i$ will not send another full map to $r_j$ until at least $\tau$ steps have elapsed. Between full exchanges, robots continue to share lightweight messages containing only location and task information, enabling fast coordination with minimal bandwidth.

Since each robot builds its map in its own local coordinate frame, merging maps requires estimating the rigid transformation between frames. We identify candidate point matches between the two maps using two methods: ORB feature matching on the obstacle grid restricted to explored regions, and semantic landmark matching that pairs object centroids of the same category by comparing the local obstacle structure around each centroid using a rotation-invariant radial descriptor. Because candidate matches inevitably include incorrect correspondences, recovering the transformation requires a robust estimator. Among approaches for occupancy grid alignment, spectral methods \cite{carpin2008} assume environments with prominent linear features. ICP \cite{besl1992icp} requires a good initial alignment and is sensitive to partial overlap. Feature-based pipelines pair a keypoint detector with a robust estimator such as RANSAC \cite{fischler1981ransac} or MSAC~\cite{torr2000mlesac} to reject outliers \cite{sensors2023feature_map_merging}. Within this family, we evaluated several detector-estimator combinations, including KAZE/MSAC \cite{sensors2023feature_map_merging} and AKAZE/RANSAC \cite{zhang2022akaze}, and found ORB/RANSAC most reliable in our setting, particularly under the limited map overlap that arises in our setup. Since the alignment stage of our pipeline is designed to be modular, the ORB/RANSAC combination can be replaced with alternative methods without affecting the rest of the system.

The alignment is validated by intersection-over-union of obstacle cells in the overlapping explored regions, and rejected below a confidence threshold.  If alignment fails, robots continue operating independently with their local map. Later encounters may succeed as additional explored regions provide more feature correspondences. Once estimated, the transformation is cached per neighbor pair to accelerate future map merges and enable the lightweight location-only exchanges described above.

During communication, robots can exchange:
\begin{itemize}
    \item \textbf{Semantic map}: The multi-layer map including occupancy, explored regions, and semantic channels. The neighbor's map is warped into the receiving robot's frame using the estimated transformation. Since both occupancy and semantic channels are stored as log-odds, merging selects the value with the greater absolute magnitude per cell, preserving the stronger evidence from either robot. Values are not summed to avoid double-counting evidence when both robots have observed the same region.
    \item \textbf{Location}: The robot's current position in its own frame, which the receiver transforms into its frame using the cached transformation.
    \item \textbf{Goal status}: Which goals have been completed, allowing other robots to remove them from their task list.
    \item \textbf{Intent}: The goal each robot is currently pursuing, used for cooperative task allocation as described below.
\end{itemize}

The communication protocol requires no central component or global connectivity; each robot builds its own understanding of the team state by accumulating information from pairwise exchanges. When a robot broadcasts its intent to pursue a goal, other robots remove that goal from their planning. To resolve conflicts when multiple robots simultaneously target the same goal, if a robot receives an intent for a goal it is currently pursuing from a robot with a higher score, it yields and selects an alternative goal. Ties are broken by a unique priority assigned at initialization. This mechanism reduces duplicate effort without negotiation.

\subsubsection{Frontier Selection}
\begin{figure}
    \centering
    \includegraphics[width=0.7\linewidth]{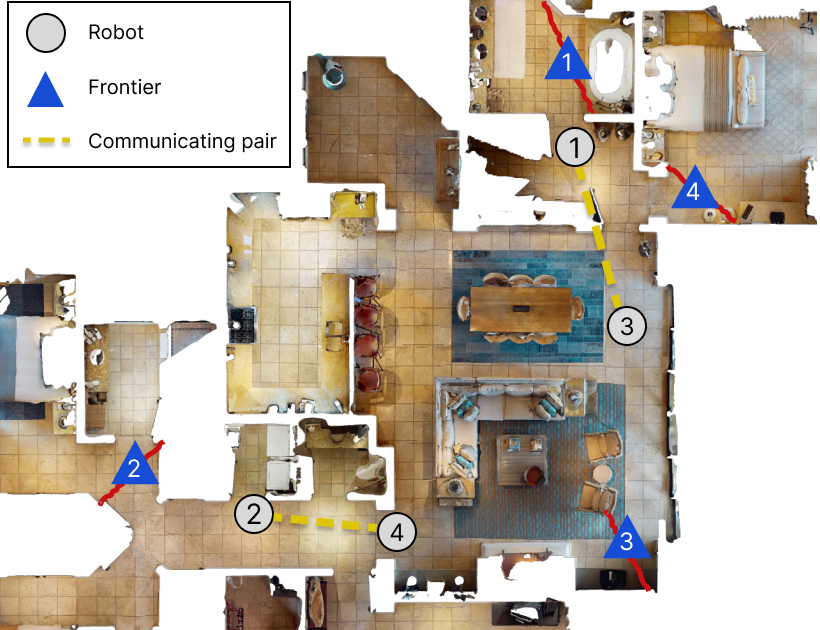}
    \caption{Frontier selection among four robots with local communication. Connected pairs (1--3 and 2--4, linked by dashed lines) spread apart to maximize coverage. Robot 4, currently lacking communication with Robots 1 and 3, targets a frontier in an area already explored by them, illustrating the challenge of achieving global coordination in a decentralized manner. This redundancy resolves once Robot 4 exchanges information with Robot 1 or 3.}
    \vspace{-12pt}
    \label{fig:comms}
\end{figure}

During exploration, robots must select the frontier that maximizes coverage while minimizing overlap with other agents. We achieve this through a distance-based weight for each frontier that incorporates neighboring robots' positions. For each candidate frontier $f$, we calculate 
\begin{equation}
    w(f, i) = \frac{\min_{j \in \calN(i)} d(f, p_j)}{d(f, p_i)},
\end{equation}
where $d(f, p_i)$ is the geodesic distance from robot $i$ at $p_i$ to the nearest cell in frontier $f$, and $\calN(i)$ is the set of neighbors of robot $i$. This formulation encourages robots to select frontiers where they are closer than others, distributing the team across the environment without explicit negotiation. \figref{fig:comms} illustrates this behavior: robots that have recently communicated select spatially separated frontiers, while Robot 4, lacking communication with Robots 1 and 3, targets a region they have already explored. This redundancy resolves once Robot 4 exchanges information with Robot 1 or 3.

Several methods exist for coordinating frontier selection in multi-robot exploration, including auction-based allocation~\cite{gerkey2002sold}, distributed variants of the Hungarian method~\cite{chopra2017distributed}, and heuristics such as MinPos~\cite{faigl2014comparison}. These require coordination rounds, connected communication, or synchronous position knowledge, none of which hold under intermittent ad-hoc connectivity. Our distance-ratio heuristic avoids these issues: each robot evaluates frontiers independently using only the most recently received neighbor positions, without requiring any coordination rounds. This makes the approach particularly suitable for our decentralized, asynchronous setting, though it could be replaced by more sophisticated allocation methods if reliable communication is available.

\section{Evaluation}
\label{sec:eval}
We evaluate our method in Habitat simulator \cite{habitat2019, habitat2021} using HM3DSem scenes. Robots operate with a discrete action space: \textit{move\_forward} (25\,cm), \textit{turn\_left/right} ($30^\circ$), and \textit{stop}. A goal is considered found when a robot executes \textit{stop} within a dataset-specified distance (0.2\,m -- 1\,m) of the target. We use a proximity-based communication model where robots can communicate within range $r_{\text{comm}} = 5\,\text{m}$.

\subsection{Dataset}
We employ two task datasets to benchmark our method.

\textbf{HM3Dv0.2:} We use the HM3Dv0.2 validation split to compare against existing multi-agent semantic navigation methods. This dataset consists of 1000 single-target categorical goal episodes across 36 scenes, limited to 6 object categories: chair, bed, couch, potted plant, toilet, and TV. A goal is considered found if the agent stops within 0.2\,m of the target. For comparison with the baselines on this dataset, we use two robots to match the configuration used in prior work \cite{conav2023}, \cite{mcoconav2025}. We use MaskR-CNN \cite{he2017mask} for object detection to ensure fair comparison with baselines that use similar closed-vocabulary detectors.

\textbf{GOAT-Bench.} While HM3Dv0.2 enables comparison with existing multi-agent methods, it lacks open-vocabulary multimodal goals. To evaluate multimodal, multi-object performance, we use GOAT-Bench \cite{goatbench2024}, which provides open-vocabulary episodes with category, language, and image goals. The dataset spans 312 object categories, and each episode contains 5--10 goals. We evaluate on the \textit{val\_seen} split, which uses the same 36 scenes as HM3Dv0.2 with 10 episodes per scene. A goal is considered found if the agent stops within 1\,m of the target (following GOAT-Bench evaluations). We use a communication cooldown period of $\tau = 10$ steps and evaluate with 1--4 agents. For the GOAT-Bench dataset, we use Detic \cite{detic2022} to support open-vocabulary detection.

\subsection{Metrics}
\label{subsec:eval_metrics}

For single-goal episodes, we report Success Rate (SR), the proportion of episodes where the agent successfully locates the target, and Success weighted by Path Length (SPL), which additionally penalizes inefficient paths. SPL is defined as \( \text{SPL} = S \cdot \frac{d^*}{\max(d, d^*)} \), where
$S \in \{0,1\}$ indicates success, $d^*$ is the length of the shortest
navigable path from the agent's starting position to the target, and $d$ is the total distance traveled by the agent.

However, SPL is designed for single-agent, single-object episodes and does not extend to our setting, with multiple robots, objects, and valid instances per object. A goal such as ``chair'' may be satisfied by any of several instances, so the optimal path depends on which instance each robot targets. This structure corresponds to the Generalized TSP (GTSP) \cite{gtsp_laporte}, where each goal defines a cluster of valid locations and visiting any node satisfies it. We extend GTSP to the multi-agent setting with a min-max objective over robot path lengths, since makespan reflects completion time under parallel execution. Moreover, robots follow open paths, not closed tours.

Based on this formulation, we propose \textbf{Multi-object Multi-agent SPL (MSPL)}, a metric that evaluates navigation efficiency for any combination of team sizes and goal counts. For multi-object episodes, SR is defined as the proportion of goals successfully located across all episodes (also known as ``Progress'' \cite{multion2020}). MSPL extends SPL as:
\begin{equation}
    \text{MSPL} = \text{SR} \cdot \frac{d^{*}}{\max \{d^*, \max_j d_j\}}
\end{equation}
where $d^{*}$ is the optimal makespan and $\max_j d_j$ is the longest distance traveled by any robot during execution. The $\max$ in the denominator ensures MSPL remains bounded in $[0, 1]$, consistent with single-agent SPL. Notably, MSPL reduces to standard SPL when $n=1$ and $m=1$ (single agent, single goal), making it a strict generalization.

We formulate the optimal makespan computation as a Mixed Integer Linear Program (MILP) and solve it using Gurobi \cite{gurobi}. Since $d^*$ depends only on the scene geometry and goal locations, it can be computed offline once per episode and does not affect runtime performance.

\subsection{Baselines}

\textbf{Co-NavGPT} \cite{conav2023} uses a Vision Language Model (VLM) as a centralized global planner. Each robot builds a local semantic map, which is aggregated into a unified global map. The VLM receives this map along with encoded robot states and frontier regions, and assigns frontiers to each robot for coordinated exploration. This requires centralized map aggregation and a single VLM for all robots.

\textbf{MCoCo-Nav} \cite{mcoconav2025} employs multimodal chain-of-thought reasoning with VLMs to evaluate exploration value at each frontier through probabilistic scoring. Each robot runs its own VLM for decentralized planning. However, all robots share a global semantic map that integrates observations from the entire team, creating implicit coupling among agents.

Both baselines handle only single categorical goals and rely on shared global state, whereas our method is fully decentralized and supports multimodal, multi-object episodes.

\begin{table}[tb]
\footnotesize
\centering
\caption{Single-agent comparison on GOAT-Bench multi-object episodes.}
\label{tab:goat_single_comparison}
\begin{tabular}{l|cc|cc}
\toprule
& \multicolumn{2}{c|}{\textbf{Predicted Sem.}} & \multicolumn{2}{c}{\textbf{Ground-Truth Sem.}} \\
\cmidrule(lr){2-3} \cmidrule(lr){4-5}
\multicolumn{1}{c|}{\textbf{Method}} & SR $\uparrow$ & SPL $\uparrow$ & SR $\uparrow$ & SPL $\uparrow$ \\
\midrule
GOAT \cite{goat2023} & 26.3 & 17.5 & 56.7 & 40.3 \\
\textbf{\texorpdfstring{DM$^3$-Nav}{DM3-Nav} (Ours)} & \textbf{44.0} & \textbf{30.7} & \textbf{85.7} & \textbf{61.6} \\
\bottomrule
\end{tabular}
\vspace{-10pt}
\end{table}

\begin{table*}[b]
\centering
\caption{Comparison on HM3Dv0.2 dataset with two agents using predicted and ground-truth semantics.}
\label{tab:hm3d_comparison}
\footnotesize

\begin{tabular*}{\textwidth}{@{\extracolsep{\fill}} l|cc|cc|cc|cc @{}}
\toprule
\multicolumn{1}{c|}{\multirow{2}{*}{\textbf{Method}}} & \multirow{2}{*}{\textbf{Map}} & \multirow{2}{*}{\textbf{Planner}} & \multirow{2}{*}{\textbf{Goals}} & \multirow{2}{*}{\textbf{Multi-Object}} & \multicolumn{2}{c|}{\textbf{Predicted Semantics}} & \multicolumn{2}{c}{\textbf{Ground-Truth Semantics}} \\
\cmidrule(lr){6-7} \cmidrule(lr){8-9}
& & & & & \textbf{SR}$\uparrow$ & \textbf{SPL}$\uparrow$ & \textbf{SR}$\uparrow$ & \textbf{SPL}$\uparrow$ \\
\midrule
Co-NavGPT & Centralized & Centralized & Category & $\times$ & 66.1 & 33.1 & 75.7 & 44.8 \\
MCoCo-Nav & Centralized & Decentralized & Category & $\times$ & 71.6 & \textbf{38.7} & 87.2 & 46.4 \\
\textbf{\texorpdfstring{DM$^3$-Nav}{DM3-Nav} (Ours)} & Decentralized & Decentralized & Category, Image, Text & \checkmark & \textbf{74.6} & 38.2 & \textbf{89.9} & \textbf{52.8} \\
\bottomrule
\end{tabular*}
\vspace{-10pt}
\end{table*}

\subsection{Results}
\label{sec:results}
We first validate our single-agent improvements on GOAT-Bench, then compare our multi-agent system against existing baselines on HM3Dv0.2, and finally evaluate the complete system on multi-agent multi-object episodes.

\subsubsection{Single-Agent Multi-Object Performance}
\label{subsec:eval:single_goat}

Table~\ref{tab:goat_single_comparison} compares our single-agent implementation against the original GOAT system on GOAT-Bench multi-object episodes with a 500-step limit. We report GOAT's published results from \cite{goatbench2024} which were evaluated on the same benchmark split. With predicted semantics, we achieve 67\% and 75\% relative improvements in SR and SPL respectively; with ground-truth semantics, improvements reach 51\% and 53\%. These significant improvements stem from our redesigned navigation components detailed in Appendix~\ref{appendix:modifications}, and they substantially strengthen the underlying semantic navigation backbone on which our decentralized framework is built.

\subsubsection{Comparison with Multi-Agent Baselines}
We compare against existing multi-agent semantic navigation approaches on the HM3Dv0.2 dataset, limiting episodes to 500 steps. As shown in Table \ref{tab:hm3d_comparison}, our method achieves the highest success rate (74.6\%) while operating fully decentralized. The slightly lower SPL compared to MCoCo-Nav (38.2 vs 38.7) reflects the inherent trade-off of decentralized coordination: without access to a shared global map, robots may occasionally explore overlapping regions before exchanging information, resulting in longer paths. However, this overhead is modest, and our method still finds more goals overall.

To highlight the impact of our coordination mechanism, we conduct two
additional experiments. First, we replace predicted semantics with
ground-truth semantics to isolate coordination from perception errors
(Table~\ref{tab:hm3d_comparison}, Ground-Truth Semantics). Our method outperforms both baselines on all metrics, showing that our decentralized approach can exceed baselines that rely on a centralized map when perception is reliable.

Second, we isolate the coordination mechanism by setting the communication range so that robots effectively never communicate ($r_{\text{comm}} = 0.1\,\text{m}$). As shown in
Table~\ref{tab:comm_ablation}, communication improves both success rate (68.6\% to 74.6\%) and SPL (28.9 to 38.2), with the greater effect on SPL because sharing maps and broadcasting intent allows robots to avoid re-exploring areas a teammate has already covered, which shortens paths. 

\begin{table}[t]
\footnotesize
\centering
\caption{Effect of communication on DM$^3$-Nav (HM3Dv0.2, two agents, predicted semantics).}
\label{tab:comm_ablation}
\begin{tabular}{l|cc}
\toprule
\textbf{Setting} & \textbf{SR} $\uparrow$ & \textbf{SPL} $\uparrow$ \\
\midrule
DM$^3$-Nav with comm. & \textbf{74.6} & \textbf{38.2} \\
DM$^3$-Nav w/o comm. & 68.6 & 28.9 \\
\bottomrule
\end{tabular}
\vspace{-10pt}
\end{table}

\begin{table}[t]
\footnotesize
\centering
\caption{Performance of DM$^3$-Nav on GOAT-Bench multi-object episodes with increasing number of robots.}
\label{tab:goat_multiagent}
\begin{tabular}{c|ccc}
\toprule
\textbf{Num. Agents} & \textbf{SR} $\uparrow$  & \textbf{Avg. Timesteps} $\downarrow$ & \textbf{MSPL} $\uparrow$ \\
\midrule
1 & 24.6 & 191.9 & 0.42 \\
2 & 30.6 & 180.2 & 0.39 \\
3 & 33.5& 173.3 & 0.35 \\
4 & \textbf{34.5} & \textbf{169.3} & 0.35 \\
\bottomrule
\end{tabular}
\vspace{-10pt}
\end{table}

\subsubsection{Multi-Agent Multi-Object Performance}

Finally, we evaluate our complete system, decentralized, multi-agent, multimodal, and multi-object navigation, on GOAT-Bench. To the best of our knowledge, no existing method supports this combination of capabilities. Therefore, we compare against our single-agent variant to quantify the benefit of multi-robot deployment.

We limit episodes to 200 total steps rather than allowing 500 steps per goal; with up to 10 goals per episode, a single robot given 5000 steps could exhaustively explore most environments, negating the benefit of parallelization. As shown in Table~\ref{tab:goat_multiagent}, a team of four robots achieves a 40\% relative improvement in success rate over a single robot (34.5\% vs 24.6\%) while also completing episodes faster. While MSPL decreases with more agents (e.g., 0.35 for four robots vs 0.42 for one), this reflects the increased coordination challenge rather than degraded performance—the optimal baseline assumes perfect task allocation, which becomes harder to approximate as team size grows. Importantly, the absolute gains in success rate and reduction in episode time demonstrate that adding robots provides clear practical benefit despite this coordination overhead.

\begin{figure}[tp]
    \centering
    \begin{subfigure}[b]{0.210\linewidth}
        \centering
        \includegraphics[width=\linewidth]{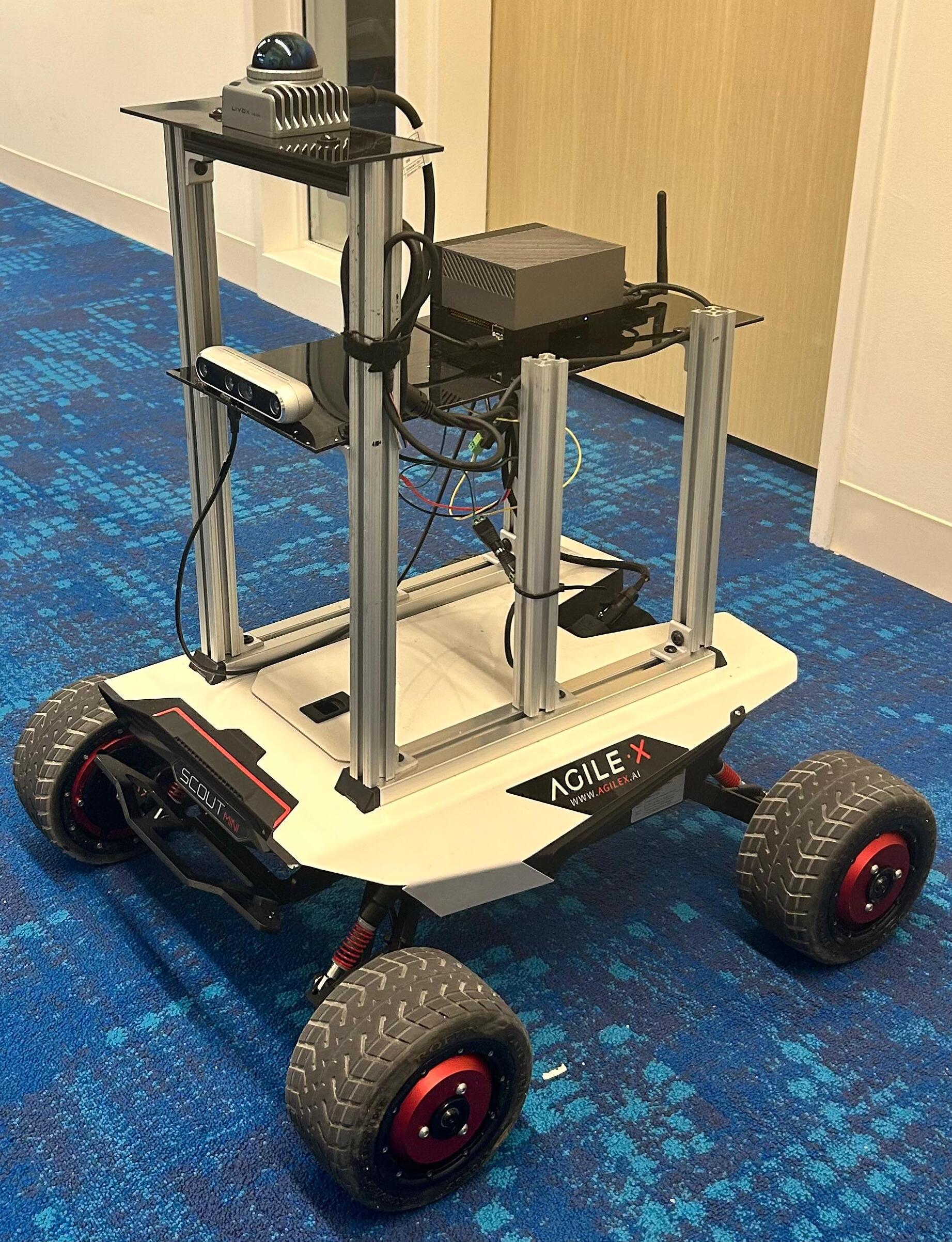}
        \caption{}
        \label{fig:agileX}
    \end{subfigure}
    \begin{subfigure}[b]{0.267\linewidth}
        \centering
        \includegraphics[width=\linewidth]{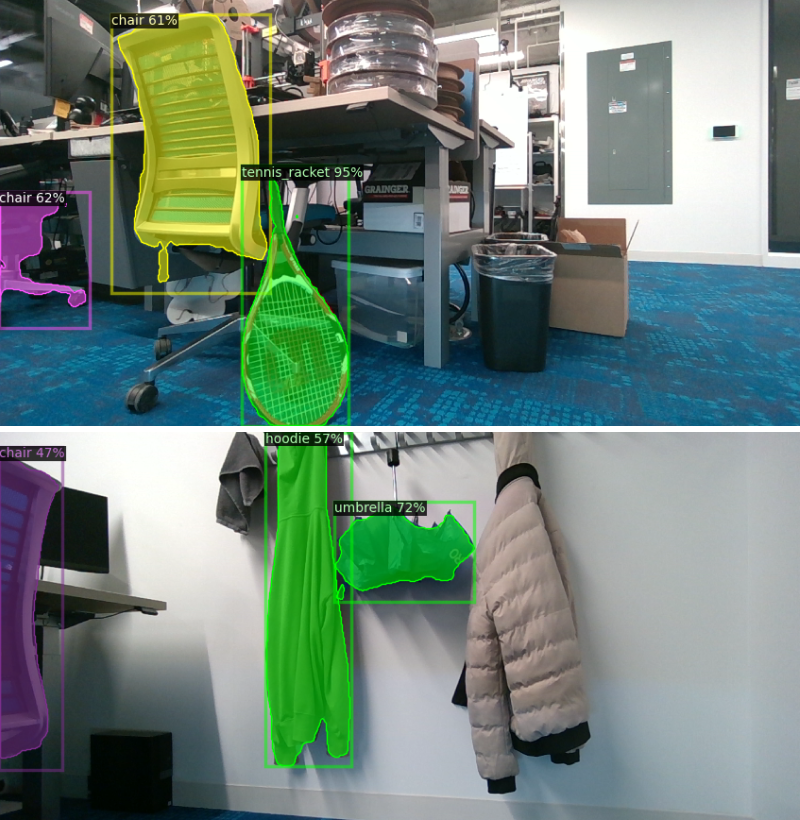}
        \caption{}
        \label{fig:sample_obs}
    \end{subfigure}
    \begin{subfigure}[b]{0.42\linewidth}
        \centering
        \includegraphics[width=\linewidth]{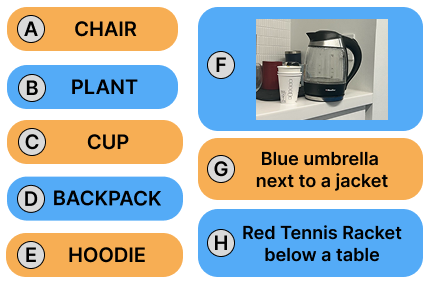}
        \caption{}
        \label{fig:goals}
    \end{subfigure}

    \caption{(a) AgileX Scout Mini with an NVIDIA Jetson Orin, RealSense D455, and a Livox MID-360 LiDAR. (b) Onboard observation and object detection. (c) Eight distinct targets to be found by the mobile robots in the environment.}
    \label{fig:main_grouped} 
    \vspace{-10pt}
\end{figure}



\begin{figure*}[t]
    \centering
    \includegraphics[width=0.9\textwidth]{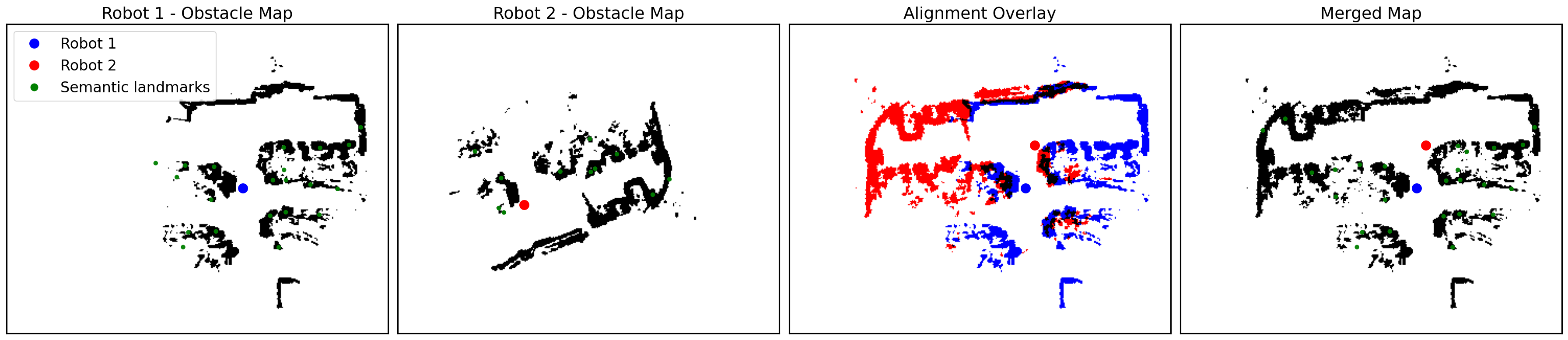}
    \caption{Map merging visualization. First two panels show the obstacle maps of Robot 1 and Robot 2 in their respective local frames before communication, with green dots indicating detected semantic landmarks used for alignment. Third panel shows the aligned overlay, with Robot 1's obstacles in blue and Robot 2's in red; overlapping regions appear in dark. Fourth panel shows the final merged map combining information from both robots.}
    \label{fig:merge}
    \vspace{-10pt}
\end{figure*}

\section{Experiments}
\label{sec:exp}
To demonstrate feasibility beyond simulation, we deploy our system on two AgileX Scout Mini mobile robots, each equipped with an NVIDIA Jetson Orin for onboard computation. Robots perceive their environment using an Intel RealSense D455 RGB-D camera for semantic mapping and a Livox MID-360 LiDAR for localization via Direct LiDAR-Inertial Odometry \cite{directlidar2022}. Consistent with our decentralized design, robots communicate through an ad-hoc WiFi network with no central infrastructure.

We evaluate our approach in an office environment of approximately 200\,$\text{m}^2$. The robots are tasked with locating 8 targets specified through all three supported modalities: category labels (backpack, hoodie, cup, chair, plant), language descriptions (tennis racket, umbrella), and a reference image (kettle), as shown in \figref{fig:goals}. \figref{fig:agileX} shows the robot platform, and \figref{fig:sample_obs} presents sample instances where the robots find a target. The two-robot team successfully located all the targets in 14 minutes. \figref{fig:merge} visualizes the map merging process: after exchanging maps with a neighboring agent, the robot incorporates its teammate's observations and location, then prioritizes exploring frontiers that maximize spatial separation.

\section{Conclusion}
\label{sec:conclusion}
We presented \texorpdfstring{DM$^3$-Nav}{DM3-Nav}, a decentralized multi-agent semantic navigation system that supports multimodal goal specifications and multi-object episodes without prior training. Our approach achieves coordination through ad-hoc pairwise communication, eliminating the need for centralized planning or global map synchronization. Experiments on HM3Dv0.2 and GOAT-Bench demonstrate that decentralized coordination matches or exceeds centralized baselines while providing architectural benefits in scalability and fault tolerance. We further validated the system's real-world applicability by deploying it on a team of mobile robots in an office environment, relying entirely on onboard sensing, localization, and computation.

Building upon this foundation, future work will pursue several directions. First, we aim to optimize the system for edge deployment by integrating lightweight perception models to reduce computational overhead and accelerate the decision-making loop. Second, we plan to extend our framework to heterogeneous robot teams, such as combining UAVs and UGVs with complementary sensing capabilities or assigning specialized roles to different agents. Third, we intend to support more complex mission specifications through temporal logics, enabling constraints on goal ordering, timing, and conditional dependencies beyond the current unordered goal sets.

\bibliographystyle{IEEEtran}
\bibliography{references}

\appendices
\section{Modifications to Single-Agent Architecture}
\label{appendix:modifications}

Our single-agent architecture follows the GOAT framework \cite{goat2023} with several implementation differences. 
Table~\ref{tab:goat_single_comparison} shows that our implementation achieves 44.0\% SR with predicted semantics and 85.7\% SR with ground-truth semantics on GOAT-Bench, compared to 26.3\% and 56.7\% for the original GOAT. We document the key differences below.


\subsection{Probabilistic Map Accumulation}
The original GOAT implementation updates its spatial map by taking the element-wise maximum between the existing map and each new observation. While this preserves detections over time, it also means that a single false positive, whether from a noisy depth reading or an incorrect segmentation, permanently marks a cell as occupied or belonging to a semantic class, with no mechanism for correction.
We replace this with a log-odds accumulation scheme for both occupancy and semantic channels. Each observation contributes positive evidence for detected occupancy or semantic class and negative evidence for free space, accumulated over time and clamped to a maximum magnitude. Cells are classified as occupied only when accumulated log-odds exceed zero, so isolated false detections decay and are eventually overridden by consistent free-space observations. This representation also integrates naturally with our multi-robot map merging (Section~\ref{subsec:dec_coord}), where we select the value with greater absolute magnitude rather than summing, to avoid double-counting evidence from overlapping observations.

\subsection{Goal Region Selection}

When navigating to a detected object, the original implementation selects the closest free cell to the object's location. This approach can fail in several scenarios: the closest free cell may be unreachable, or it might be on the opposite side of a wall. 
We improve goal region selection through a three-step process: (1) dilate the goal region to identify nearby free cells, (2) cluster these free cells into connected components, and (3) select the cluster closest to the agent's current viewpoint. This ensures that the selected goal position is both reachable and observable.

\subsection{Frontier-Based Exploration}
The original GOAT implementation marks explored regions using geometric heuristics, such as placing a fixed-radius circle around the agent's position. This does not account for the agent's actual field of view and can mark areas outside walls as explored.
We replace this with a raycasting-based approach that marks only cells visible from the agent's current position, accounting for field of view and obstacle occlusions. This provides an accurate representation of what the agent has actually observed.

\subsection{Local Planning}

We make several improvements to low-level path execution. The original implementation selects waypoints from free cells within a fixed radius around the agent. We extend this by: (1) dilating obstacles when possible to maintain clearance from walls, (2)  filtering cells that are free but blocked along the path, and (3) adaptively reducing the planning radius when larger radii yield no valid waypoints, enabling finer movements in constrained spaces.

\subsection{Detection Verification}

To reduce false positive detections, we add a verification stage using YOLO as a secondary detector. For an object detected by the primary detector (MaskR-CNN or Detic), we also compute a YOLO detection score and average it with the initial score when deciding whether to navigate to the object. We use YOLOv10 \cite{yolov10_2024} for closed-vocabulary detection on HM3Dv0.2 and YOLO-World \cite{YOLOWorld2024} for open-vocabulary detection on GOAT-Bench. This verification step improves the success rate by filtering out incorrect detections.

\subsection{Additional Modifications}

\textbf{Stair detection.} We detect stairs by identifying regions where expected ground plane points are missing in the point cloud, and mark these as non-navigable.

\textbf{Map scale switching.} When no valid path exists in the local map, we switch to planning on the global map. 

\textbf{Detection score thresholding.} For categorical goals, rather than navigating to the highest-scoring detection immediately, we consider the detection confidence and navigate only when it exceeds a threshold.

\textbf{Segmentation refinement.} We apply morphological erosion to segmentation masks to reduce noise, and filter out points whose depth deviates significantly from the instance median to remove incorrectly included background regions.

\end{document}